\documentclass[slac_one]{revtex4}
\usepackage{graphicx}
\usepackage{fancyhdr}
\newcommand{\etal}{\it et al.\rm}

\begin{document}

\title{Light Hadron Spectroscopy and Charmonium} 

%

\author{Frederick A. Harris (from the BES Collab.)}
\affiliation{Dept. of Physics and Astronomy, University of Hawaii,
  Honolulu, Hawaii 96822, USA}

\begin{abstract}
During the last few years there has been a renaissance in charm and
charmonium spectroscopy with higher precision measurements at the
$\psi^{'}$ and $\psi(3770)$ coming from BESII and CLEOc and many new
discoveries coming from B-factories.  In this paper, I review some new results
on ``classical'' charmonium and $e^+ e^- \to$ hadrons using B-factory Initial
State Radiation and two photon events.

\end{abstract}

\maketitle

\thispagestyle{fancy}


\section{Introduction}

In recent years, there has been tremendous progress in charm and
charmonium spectroscopy with many new results from Belle, BaBar, CDF,
D0, BES and CLEOc. I will review recent results on ``classical''
charmonium; Galina Pakhlova (see her paper in these proceedings)
covers exotic charmonium states: X, Y, Z, etc.  

Charmonium provides detailed information on QCD in the perturbative
and non-perturbative regimes, as well as providing a laboratory for
precision tests of lattice QCD and effective field
theory~\cite{rosner}.

Charmonium may be produced by $e^+e^-$ annihilation, $B$ decays,
$e^+e^-$ annihilation where one or both electrons loses energy by
Initial State Radiation (ISR), and two photon processes.  The very
high luminosity of B-factories allow the use of the latter two even though
their cross sections are suppressed.  Only $J^{PC} = 1^{--}$ charmonium
states can be produced directly in $e^+e^-$ annihilation.  However
states below the $\psi(2S)$ may be produced by $\psi(2S)$ radiative
and hadronic transitions, a technique which has been used extensively
by BES and recently by CLEOc.

I will also cover some light hadron spectroscopy results, including
the $Y(2175)$ and $\phi(1680)$, two photon results at Belle, and ISR
results at BaBar, and I will report the status of BEPCII and BESIII.
I apologize to everyone whose results I do not cover because of lack
of time and space.  Please see the references for details.

\section{Charmonium results}

Recently CLEOc obtained 27 million $\psi(2S)$ events and has reported
many new high precision results using this sample. This is the world's
largest $\psi(2S)$ sample produced in $e^+e^-$ collisions.  

\subsection{\boldmath $\psi(2S) \to J/\psi$ transitions}
CLEOc has studied $\psi(2S) \to J/\psi$ transitions using $J/\psi \to
e^+e^-$ and $\mu^+\mu^-$ decays to identify the
$J/\psi$~\cite{psi2stojpsi}.  A summary of their branching fraction
results is shown in Table~\ref{psi2s2j}, along with those of the
Particle Data Group (PDG).  Using the product $\chi_c$ branching
fractions and $B(\psi(2S) \to \gamma \chi_{cJ}) = (9.4 \pm 0.4)\%,$
$(8.8\pm0.04)\%,$ and $(8.3 \pm 0.4)\%$~\cite{PDG08} for $J=0,1,2$,
respectively, the $B(\chi_{cJ} \to \gamma J/\psi)$ radiative branching
fractions are obtained and also listed in Table~\ref{psi2s2j}.
Precision branching fractions are important since $\psi(2S)$
production and decay is a primary means of producing $\chi_c$ states,
and $\psi(2S)$ production in many experiments is measured by the
narrow mass peak recoiling against the $\pi^+\pi^-$ in $\psi(2S) \to
\pi^+ \pi^- J/\psi$ decay.

\begin{table}[htb]
\begin{center}
\caption{\label{psi2s2j} Branching fractions for $\psi(2S) \to
J/\psi$ transitions from CLEOc~\cite{psi2stojpsi}.  $B(\psi(2S) \to
\gamma \chi_{cJ})$ values from Ref.~\cite{PDG08} are used to obtain
the CLEOc $\chi_c \to \gamma J/\psi$ radiative branching fractions,
and the last error for them is from this branching fraction.}
\begin{tabular}{|l|c|c|}
\hline Channel & B(\%) CLEOc & B(\%) PDG08~\cite{PDG08} \\ \hline
$\pi^+\pi^- J/\psi$ & $35.04 \pm 0.07 \pm 0.77$ & $32.6\pm0.5$ \\
$\pi^o\pi^o J/\psi$ & $17.69 \pm 0.08 \pm 0.53$ & $16.84\pm0.33$ \\
$\eta J/\psi$ & $3.43 \pm 0.04 \pm 0.09$ & $3.16\pm0.07$ \\
$\pi^o J/\psi$ & $0.133 \pm 0.008 \pm 0.003$ & $0.126\pm0.013$ \\
$\gamma \chi_{c0} (\chi_{c0} \to J/\psi)$ & $0.125 \pm 0.007 \pm 0.013$
& $0.120\pm0.010$ \\
$\gamma \chi_{c1} (\chi_{c1} \to J/\psi)$ & $3.56 \pm 0.03 \pm 0.12$
& $3.15\pm0.08$ \\
$\gamma \chi_{c2} (\chi_{c2} \to J/\psi)$ & $1.95 \pm 0.02 \pm 0.07$ & $1.66\pm0.04$ \\
anything $J/\psi$ & $62.54 \pm 0.16 \pm 1.55$ & $57.4\pm0.9$ \\ \hline
$\chi_{c0} \to \gamma J/\psi$ & $1.32 \pm 0.07 \pm 0.14 \pm0.06$
&$1.28\pm0.11$ \\
$\chi_{c1} \to \gamma J/\psi$ & $40.5 \pm 0.3 \pm 1.4 \pm1.8$
&$36.0\pm1.9$ \\
$\chi_{c2} \to \gamma J/\psi$ & $23.5 \pm 0.2 \pm 0.8 \pm1.1$
&$20.0\pm1.0$ \\
\hline
\end{tabular}
\end{center}
\end{table}

\subsection{\boldmath $J/\psi$ and $\psi(2S) \to \gamma \eta_c$}
\label{gammaetac}
The precise determination of the $\eta_c$ mass provides information on
the hyperfine splitting of the $\eta_c$ and $J/\psi$. However,
although there have been many measurements of the $\eta_c$ mass and
width, the measurements do not agree very well, and the fitted masses
and widths in the PDG~\cite{PDG08} have very low
confidence levels: 0.002 for the mass and $< 0.0001$ for the width.
Further, the masses obtained from $J/\psi$ and $\psi(2S)$ decays are
about 5.3 MeV/c$^2$ or $3 \sigma$ lower than those obtained from
$\gamma \gamma$ fusion and $p \bar{p}$ annihilation.  Another problem
is that the branching fraction for $J/\psi \to \gamma \eta_c$,
$B(J/\psi \to \gamma \eta_c) = 1.3\pm0.4 \%$~\cite{PDG08}, is very
low compared to recent Lattice QCD results, $B(J/\psi \to \gamma
\eta_c) = 2.1\pm0.1\pm0.4 \%$~\cite{dudek}.

Studying $J/\psi \to \gamma \eta_c$ and $\psi(2S) \to \gamma \eta_c$
is very important since these are magnetic dipole (M1) transitions,
and their branching fractions are necessary for normalizing $\eta_c$
branching fractions.

\begin{figure}[htb]
\centering
\includegraphics[width=75mm]{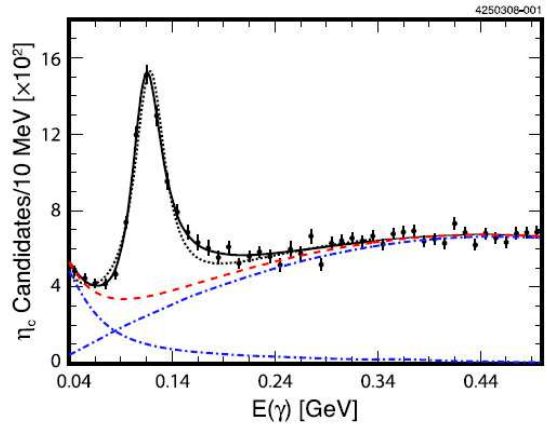}
\caption{Gamma energy spectrum from CLEOc~\cite{getac} for $J/\psi \to
\gamma \eta_c$ using $\psi(2S) \to \pi^+ \pi^- J/\psi$, $J/\psi \to
\gamma \eta_c$, $\eta_c \to X_i$ events, where $X_i$ denotes 12
exclusive $\eta_c$ hadronic decays. A fit using a Breit-Wigner plus
resolution function (dotted curve) is unable to fit the data (dots
with error bars).} \label{lineshape}
\end{figure}

CLEOc studied these decays using 24.5 million $\psi(2S)$
events~\cite{getac}.  They obtained the branching fractions by fitting
the $\gamma$ energy spectrum using three samples, (1) $\psi(2S) \to
\gamma \eta_c$ inclusive, (2) $\psi(2S) \to \pi^+ \pi^- J/\psi$,
$J/\psi \to \gamma \eta_c$, $\eta_c \to X_i$, and (3) $\psi(2S) \to
\gamma \eta_c$, $\eta_c \to X_i$, where $X_i$ denotes 12 exclusive
$\eta_c$ hadronic decays. Sample (2) was used to determine the
$\gamma$ line shape, shown in Fig.~\ref{lineshape}.  Interestingly,
they find that the line shape can not be fitted by a simple
Breit-Wigner plus a resolution function, and an empirical shape was
used. Their branching fraction results compared with the PDG are shown in
Table~\ref{etac}.  The values are very different than the PDG and will
affect all $\eta_c$ branching fractions. Their $B(J/\psi \to \gamma
\eta_c)$ agrees much better with the Lattice QCD prediction above.

\begin{table}[htb]
\begin{center}
\caption{\label{etac} Branching fraction results for $\psi(2S)$ and
$J/\psi \to \gamma \eta_c$ transitions from CLEOc~\cite{getac}.}
\begin{tabular}{|l|c|c|}
\hline Channel & CLEOc & PDG08~\cite{PDG08} \\ \hline
$\psi(2S) \to \gamma \eta_c$ & $(4.32\pm0.16\pm0.60)\times 10^{-3}$ &
$(3.0\pm0.5) \times 10^{-3}$ \\
$J/\psi \to \gamma \eta_c$ & $(1.98 \pm 0.009 \pm 0.30)\%$ & $(1.3\pm0.4)\%$ \\
\hline
\end{tabular}
\end{center}
\end{table}

If they fit the three samples with an unmodified Breit-Wigner, they
obtain a mass $m(\eta_c) = 2976.7 \pm 0.6$ MeV/c$^2$ (stat. error),
compared to $m(\eta_c) = 2982.2 \pm 0.6$ MeV/c$^2$ with their
empirical line shape. The first mass is consistent with other
determinations from $J/\psi$ and $\psi(2S)$ decay, while the second is
consistent with those coming from $\gamma \gamma$ fusion and $p
\bar{p}$ annihilation. As examples of recent $2 \gamma$ production
$m(\eta_c)$ measurements, Belle determined $m(\eta_c) = 2986.1 \pm 1.0
\pm 2.5$ MeV/c$^2$ in $\eta_c \to$ four body decays~\cite{etacfour}
and $m(\eta_c) = 2981.4 \pm 0.5 \pm 0.4$ MeV/c$^2$ using $\eta_c \to
K_S K \pi$~\cite{etackskpi}. The line shape problem may explain the $3
\sigma$ mass difference with $\gamma \gamma$ fusion and $p \bar{p}$
annihilation, but the uncertainty in how to deal with the line shape,
according to the authors, prohibits the precise determinations of the
$\eta_c$ mass and width.  The authors point out that understanding the
energy dependence of the $\psi(1S,2S) - \gamma \eta_c$ matrix element
is crucial for understanding radiative decays.

Christine Davies showed a comparison of Lattice QCD (2007
HPQCD/MILC/FNAL) results with experiment~\cite{Davies}.  One of the
biggest discrepancies was in the $m(J/\psi) - m(\eta_c)$ comparison.
Assuming from the CLEOc mass measurements with and without using the
empirical line shape that the average $\eta_c$ mass might shift
upwards by roughly 3 MeV/c$^2$, the agreement between LQCD prediction
and experiment for $m(J/\psi) - m(\eta_c)$ would be considerably
improved.

\subsection{\boldmath $h_c(^1P_1)$}

In 2005, E835~\cite{hc_e835} and CLEO~\cite{hc_cleoc} reported
measurements of the mass of the $h_c(^1P_1)$.  CLEO used $e^+ e^-
\to \psi(2S) \to \pi^0 h_c$ and determined the $h_c$ mass by measuring
the mass recoiling against the $\pi^0$ using both $h_c \to \gamma
\eta_c$ inclusive events and exclusive $ \eta_c$ decays.  They have
repeated their analysis using the 25 million $\psi(2S)$
sample~\cite{cleohc}.  They find excellent agreement between the
inclusive and exclusive results and obtain a combined result of
$m(h_c) = 3525.28 \pm 0.19 \pm 0.12$ MeV/c$^2$.  Combining with their
2005 result, they obtain $m(h_c)_{AVG} = 3525.2 \pm 0.18 \pm 0.12$
MeV/c$^2$ and a product branching fraction, $(B_1 \times B_2)_{AVG} =
(4.16 \pm 0.30 \pm 0.37) \times 10^{-4}$.  A precise determination of
the mass is important to learn about the hyperfine (spin-spin)
interaction of the $P$ wave states.  Using the spin weighted centroid
of the $^3P_J$ states, $<m(^3P_J)>$, to represent $m(^3P_J)$, they
obtain $\Delta m_{hf}(1P) = <m(^3P_J)> - m(^1P_1) = +0.08 \pm 0.18 \pm
0.12$ MeV/c$^2$.  This is consistent with the lowest order expectation
of zero.

\subsection{\boldmath $\chi_{cJ} \to \gamma \gamma$}
\label{chictogg}
$\chi_{cJ}(^3P_J) \to \gamma \gamma$ decays (QCD) are analogous to triplet
decays of positronium (QED).  For $R= \Gamma(^3P_2 \to \gamma
\gamma)/\Gamma(^3P_0 \to \gamma \gamma)$, even the differences due to
different masses and wave functions cancel, and for both
$R = 4/15 \sim 0.27$~\cite{novikov}.
Departures from this are from strong radiative corrections and
relativistic effects.  

\begin{figure}[htb]
\centering
\includegraphics[width=75mm]{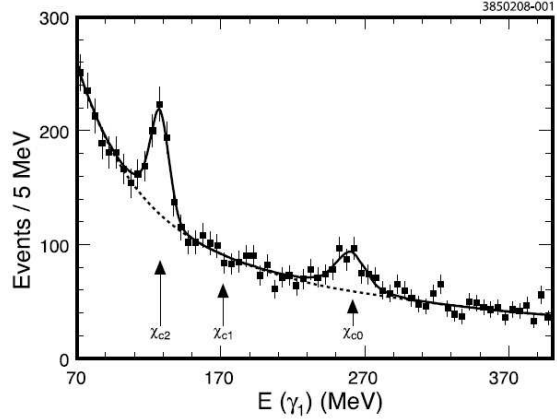}
\caption{Energy spectrum of $\gamma_1$ in $\psi(2S) \to \gamma_1 \chi_{cJ},
\chi_{cJ} \to \gamma_2 \gamma_3$ from CLEOc~\cite{cleogg}.} \label{fig:chicjtogg}
\end{figure}

CLEOc has studied this process using $\psi(2S) \to \gamma_1 \chi_{cJ},
\chi_{cJ} \to \gamma_2 \gamma_3$ with their 25 million event
sample~\cite{cleogg}.  The $\gamma_1$ energy spectrum is shown in
Fig.~\ref{fig:chicjtogg}, where clear peaks corresponding to
$\chi_{c0}$ and $\chi_{c2}$ are seen.  $\chi_{c1}$ is forbidden by the
Landau-Yang Theorem~\cite{landauyang}.  Results are listed in
Table~\ref{chicjtogg}.  Averaging the value of $R$ with those from
previous experiments, $<R> = 0.20 \pm 0.02$ is obtained.  The
theoretical first order pQCD prediction for $R$ is $R_{Th} =
(4/15)[1-1.76\alpha_S]$~\cite{barbieri}, and for $\alpha_S = 0.32$,
$$R_{Th} = 0.12.$$ The
disagreement with the experimental result confirms the inadequacy of
the first order radiative corrections.

Belle has also determined $R$ using a number of exclusive decays of
the $\chi_c$'s produced in $\gamma \gamma \to
\chi_{c0,2}$~\cite{etacfour}. They measure $\Gamma_{\gamma \gamma} \times
B(\chi_{c0,2} \to X)$, where $X$ is an exclusive decay,  and divide by
the known branching fraction to
obtain $\Gamma_{\gamma \gamma}$.  The results are also listed in
Table~\ref{chicjtogg}, and their $R$ values agree very well with
$<R>$ determined by CLEOc.

\begin{table}[htb]
\begin{center}
\caption{\label{chicjtogg} $\chi_{cJ} \to \gamma \gamma$ widths and
  $R= \Gamma(^3P_2 \to \gamma \gamma)/\Gamma(^3P_0 \to \gamma \gamma)$
  from CLEOc~\cite{cleogg} for $\chi_{cJ} \to \gamma \gamma$ and
  Belle~\cite{etacfour} for $\gamma \gamma \to \chi_{cJ}$, with
  $\chi_{cJ}\to$ exclusive hadronic decays. The CLEOc result is the
  first result; their last error is the contribution
  from $B(\psi(2S)\to\gamma\chi_c)$~\cite{PDG08}.}
\begin{tabular}{|l|c|c|c|}
\hline Channel &  $\Gamma_{\gamma \gamma}(\chi_{c0})$ &
$\Gamma_{\gamma \gamma}(\chi_{c2})$ & $R$ \\ \hline
$\chi_{cJ} \to \gamma \gamma$ & $2.53 \pm 0.37 \pm 0.11 \pm 0.24$ keV &
$0.60 \pm 0.06 \pm 0.03 \pm 0.05$ keV & $0.237 \pm 0.043 \pm 0.015 \pm
0.031$\\ \hline
$\to K_SK_S$ & $2.53 \pm 0.23 \pm 0.40$ keV & $0.46 \pm 0.08 \pm 0.09$ keV
& $0.18 \pm 0.03 \pm 0.04$ \\
$\to 4 \pi$ & $1.84 \pm 0.15 \pm 0.27$ keV & $ 0.40 \pm 0.04 \pm 0.07$ keV
& $0.22 \pm 0.03 \pm 0.05$ \\
$\to 2K2\pi$ & $2.07 \pm 0.20 \pm 0.40$ keV & $ 0.44 \pm 0.04 \pm 0.16$
keV & $0.21 \pm 0.03 \pm 0.09$ \\
$\to 4K$ & $2.88 \pm 0.47 \pm 0.53$ keV & $0.62 \pm 0.12 \pm 0.12$ keV & $
0.21 \pm 0.06 \pm 0.06$ \\  
\hline
\end{tabular}
\end{center}
\end{table}

\subsection{\boldmath Anomalous line shape of $\sigma(e^+ e^- \to hadrons)$
  in the $\psi(3770)$ energy region}

BESII has accumulated more than 30 pb$^{-1}$ of data in the region of
the $\psi(3770)$ from 3.650 to 3.872 GeV, while CLEOc has 818
pb$^{-1}$ at the $\psi(3770)$.  These samples have provided precision
measurements of $D$ meson decays using the very clean $\psi(3770) \to
D \bar{D}$ events, as well as vastly improved knowledge about the
$\psi(3770)$.  However, there has been a puzzle concerning the amount of
non-$D\bar{D}$ decay of the $\psi(3770)$.  The  $\psi(3770$ is just
above threshold for $D\bar{D}$ production and is expected to decay
into  $D\bar{D}$ pairs with a branching fraction greater than 98\%.
Surprisingly, BES measured the branching fraction of $\psi(3770$ decays to
$D\bar{D}$ to be $B(\psi(3770) \to D\bar{D} =
(85\pm3)\%$~\cite{rg1,rg2,PDG08} and directly measured $B(\psi(3770) \to
{\rm non-}D\bar{D} = (13.4\pm5.0\pm3.6)\%$~\cite{rg4} and $B(\psi(3770) \to
{\rm non-}D\bar{D} = (15.1\pm5.6\pm1.8)\%$~\cite{rg5}.  However, BES and CLEOc
have searched for exclusive non-$D\bar{D}$ decays of the
$\psi(3770)$, and the summed  non-$D\bar{D}$ branching fractions
measured by each of the collaborations are less than
2\%~\cite{rg6,cleo3770}.

\begin{figure}[htb]
\centering
\includegraphics[width=75mm]{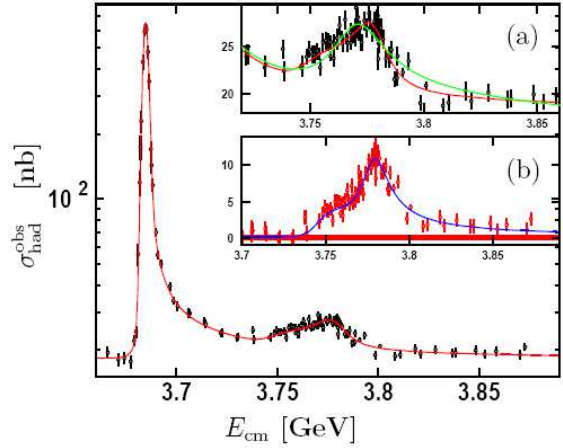}
\caption{Inclusive hadronic cross section in the region of the
  $\psi(3770)$ by BES versus center of mass (CM)
  energy~\cite{rganomalous}. (a) Fit with one resonance and (b) fit
  with two interfering resonances.}
\label{fig:anomalous}
\end{figure}

In the inclusive measurements, BES assumed a single resonance in the
energy region between 3.7 and 3.872 GeV.  To understand the
discrepancy between the inclusive and exclusive measurements, BES has
reanalyzed the fine $R$-scan ($R=\frac{e^+e^- \to hadrons}{e^+e^- \to
  \mu^+\mu^-}$) in this region and finds that the fit to
a single resonance is very poor and that allowing two non-interfering
or two interfering resonances gives a much better fit, as seen in
Fig.~\ref{fig:anomalous}~\cite{rganomalous}.  The large non-$D\bar{D}$
inclusive branching fractions measured by BES may be due partially to the
assumption of only one simple resonance in this region.  This cross
section anomaly must be confirmed, and this will be a high priority
for BEPCII and BESIII (see below).

\section{\boldmath $Y(2175)$ and $\phi(1680$)}
A structure at 2175 MeV in the $\phi f_0(980)$ mass was observed by
BaBar in the ISR process $e^+e^- \to \gamma_{_{ISR}} \phi f_0(980)$;
the mass and width are $m(Y(2175))=2175 \pm 10 \pm 15$ MeV/c$^2$ and
$\Gamma(Y(2175)) = 58 \pm 16 \pm 20$ MeV/c$^2$~\cite{babar2175}. BaBar
speculated that the $Y(2175)$ is the $s\bar{s}$ version of the
$Y(4260)$~\cite{babarspec} since it is also a $1^{--}$ and has
somewhat similar decay properties~\cite{babar2175}.

\begin{figure}[htb]
\centering
\includegraphics[width=65mm]{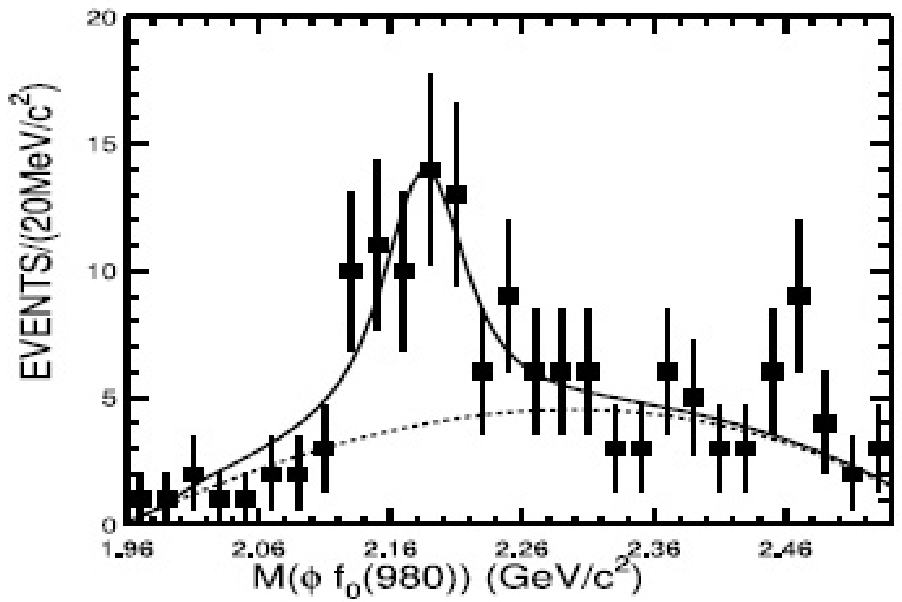}
\includegraphics[width=65mm]{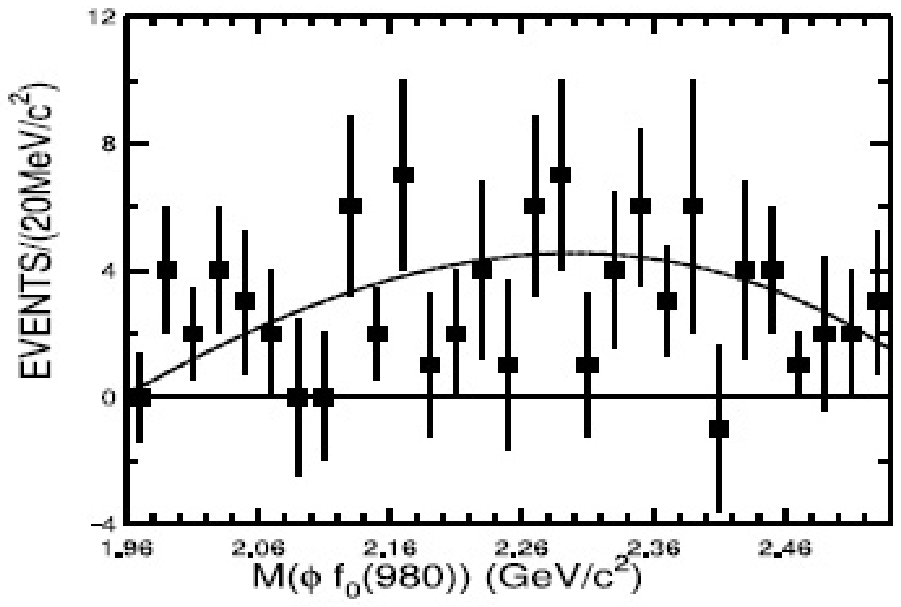}
\put(-330,96){(a)}
\put(-140,96){(b)}
\caption{Fit of $Y(2175)$ by BES~\cite{bes2175}. {\bf (a)} Fit (solid
  curve) to the data (points with error bars) versus $m(\phi
  f_0(980)$; the dashed curve indicates the background function. {\bf
  (b)} Simultaneous fit to the sideband background events (points with
  error bars) with the same background function.  The background
  normalizations for the two plots are constrained to be equal.}
  \label{fig:bes2175}
\end{figure}

BES searched for the $Y(2175)$ in $J/\psi \to \eta \phi f_0(980), \eta
\to \gamma \gamma, \phi \to K^+ K^-, f_0(980) \to \pi^+ \pi^-$ using
58 million $J/\psi$ events and found a peak in the $\phi f_0(980)$
mass around 2175 MeV/c$^2$~\cite{bes2175}.  Fig.~\ref{fig:bes2175}
shows the simultaneous fit to signal and sideband events with a
Breit-Wigner to represent the signal and a third order polynomial for
the background. The peak has a significance of about 5 $\sigma$, and
the mass and width obtained are $m(Y(2175))= 2186 \pm 10 \pm 6$
MeV/c$^2$ and $\Gamma(Y(2175)) = 65 \pm 23 \pm 17$ MeV/c$^2$, in good
agreement with BaBar.  Fitting also the smaller $\sim2 \sigma$ peak at
around 2460 MeV/c$^2$, also seen by BaBar, does not change the mass
and width of the first peak.  The product branching fraction is
$B(J/\psi \to \eta Y(2175))\cdot B(Y(2175) \to \phi f_0(980))\cdot
B(f_0(980) \to \pi^+ \pi^-) = (3.23 \pm 0.75 \pm 0.73) \times
10^{-4}$.

\begin{figure}[htb]
\centering
\includegraphics[width=150mm]{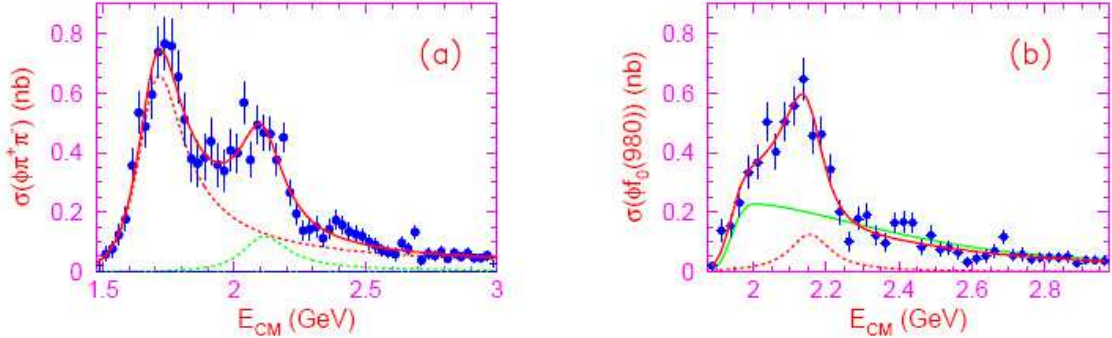}
\caption{Belle $Y(2175)$ and $\phi(1680)$ fits. (a) Fit to $\phi \pi^+
  \pi^-$ cross section in $e^+e^- \to \gamma_{_{ISR}} \phi \pi^+
  \pi^-$ and (b) fit to $\phi f_0(980)$ cross section in $e^+e^- \to
  \gamma_{_{ISR}} \phi f_0(980)$ with one Breit-Wigner (dashed curve)
  interfering with a non-resonant contribution (lower smooth
  curve)~\cite{belle2175}.}
  \label{fig:belle2175}
\end{figure}

Belle has also searched for the $Y(2175)$ in $e^+e^- \to
\gamma_{_{ISR}} \phi f_0(980)$ using 673 fb$^{-1}$ of data at the
$\Upsilon(4S)$~\cite{belle2175}.  They find a peak in the $e^+e^- \to
\phi f_0(980)$ cross section and fit it with one Breit-Wigner
interfering with a non-resident background (see
Fig.~\ref{fig:belle2175} b), as in the BaBar analysis, and also with
two Breit-Wigner functions.  They also see a peak in $e^+e^- \to
\gamma_{_{ISR}} \phi \pi^+ \pi^-$ (see Fig.~\ref{fig:belle2175} a)
and fit it.  For their final results, they take a simple average of
their fits and enlarge errors to cover the spread.  They find
$m(Y(2175)) = 2133^{+69}_{-115}$ MeV/c$^2$ and $\Gamma(Y2175)) =
169^{+105}_{-92}$ MeV/c$^2$, where the Belle errors include both
statistical and systematic errors.

Belle also finds for the first time in $e^+e^- \to \gamma_{_{ISR}}
\phi \pi^+ \pi^-$ a clear $\phi(1680)$ (see Fig.~\ref{fig:belle2175}
a), which was first seen by DM1 25 years ago~\cite{dm1}.  Belle
determines $m(\phi(1680)) = 1687\pm21$ MeV/c$^2$ and
$\Gamma(\phi(1680)) = 212\pm29$ MeV/c$^2$. BaBar also recently
reported the $\phi(1680)$ in $e^+e^- \to \gamma_{_{ISR}} \phi \eta$
and $\gamma_{_{ISR}}K^*K$~\cite{babarphi1680}.  The masses and widths
are summarized in Table~\ref{mgy2175} along with the PDG~\cite{PDG08}
$\phi(1680)$ values.  The Belle results are preliminary.

\begin{table}[htb]
\begin{center}
\caption{\label{mgy2175} Summary of masses and widths of $Y(2175)$
  and $\phi(1680)$.}
\begin{tabular}{|l|l|c|c|}
\hline Experiment & Channel &  Mass (MeV/c$^2$) & Width (MeV/c$^2$) \\ \hline
BaBar~\cite{babar2175} & $Y(2175)\to \phi f_0(980)$ & $2175\pm10\pm15$ & $58\pm16\pm20$ \\
BES~\cite{bes2175} & $Y(2175)\to \phi f_0(980)$ & $2186\pm10\pm6$ &
$65\pm23\pm17$ \\
Belle~\cite{belle2175} & $Y(2175)\to \phi \pi^+ \pi^-, \phi f_0(980)$ &
$2133^{+69}_{-115}$ & $169^{+105}_{-92}$  \\ \hline
Belle~\cite{belle2175} & $\phi(1680)\to \phi\pi^+\pi^-$ & $1687\pm 21$ & $212\pm29$ \\
BaBar~\cite{babarphi1680} & $\phi(1680)\to K^*K$ and $ \phi
\eta$  & $1709\pm20\pm43$&$322\pm77\pm160$ \\
PDG~\cite{PDG08} & $\phi(1680)$  &  $1680\pm20$ & $150\pm50$ \\
\hline
\end{tabular}
\end{center}
\end{table}

Belle finds a wider $Y(2175)$ and notes that the widths
of the $Y(2175)$ and $\phi(1680)$ are rather similar, which suggests the
possibility that the $Y(2175)$ may be an excited $\phi$ state.  So what is
the $Y(2175)$?  It could be a $s\bar{s}$ analogue of the $Y(4160)$, as
suggested by BaBar; a $s\bar{s}g$ hybrid~\cite{ding}; a $2^3D_1$
$s\bar{s}$ state~\cite{ding2}; a $s\bar{s}s\bar{s}$ tetraquark
state~\cite{wang}; a $\Lambda \bar{\Lambda}$ state~\cite{klempt}; or,
as suggested, by Belle a conventional $s\bar{s}$ state.
More data are needed to understand the $Y(2175)$.

\section{\boldmath Belle $2 \gamma$ physics}

Two $\gamma$ collisions provide valuable information on both light and
heavy quark resonances, perturbative and non-perturbative Quantum
Chromodynamics, and hadron production mechanisms. Some $\gamma\gamma$
charmonium results have already been reported in
Sections~\ref{gammaetac} and \ref{chictogg}. 
Recently Belle studied $2 \gamma$ production of $\pi^+
\pi^-$~\cite{bellepippim}, and now has new high statistics results on
$2 \gamma$ production of $\pi^0 \pi^0$ using 95 fb$^{-1}$ of
data~\cite{bellepi0pi0}.

\begin{figure}[htb]
\centering
\includegraphics[width=75mm]{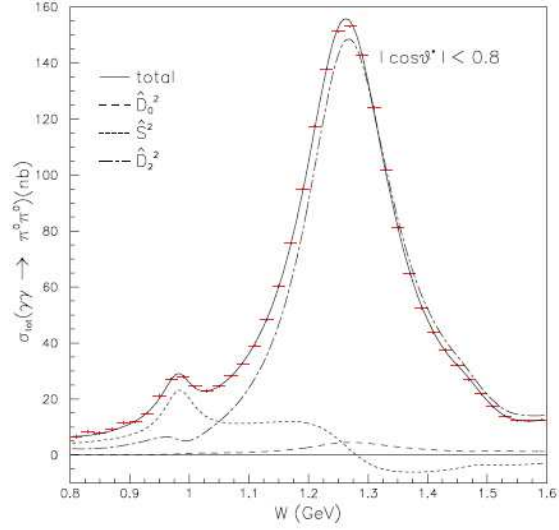}
\caption{Partial wave analysis fit of the $2 \gamma$ to $\pi^0 \pi^0$
  differential cross section in the low CM energy (W) region in terms
  of $S$, $D_0$, and $D_2$ partial waves by
  Belle~\cite{bellepi0pi0}. The contributions of the components are
  also shown.}
\label{fig:bellempi0pi0}
\end{figure}

Shown in Fig.~\ref{fig:bellempi0pi0} is a partial wave analysis fit of
the differential cross section in the low mass energy region in terms
of $S$, $D_0$, and $D_2$ partial waves.  The D$_2$ wave is dominated
by the $f_2(1270)$ while the S wave contribution includes at least one
additional resonance ($f_0(Y)$) besides the $f_0(980)$, which could be
the $f_0(1370)$ or the $f_0(1500)$.  The fit includes the $f_0(980)$,
another scalar, $f_0(Y)$, $f_2(1270)$, and $f_2^{'}(1525)$.  Known
parameters are used for the $f_2(1270)$ and the $f_2^{'}(1525)$ in the
fit.  $r_{02}$ is the ratio of helicity 0 to helicity 2 of the $f_2(1270)$.

The fit results are reported in Table~\ref{bellepi0pi0}.  The
fit with the $f_0(Y)$ included is strongly favored. BES studied
$J/\psi \to \gamma \pi \pi$ using 58 million $J/\psi$ events and found
a scalar with $m = 1466\pm6\pm20$ MeV/c$^2$ and $\Gamma =
108^{+14}_{-11} \pm 25$ MeV/c$^2$~\cite{bespipi}, in good agreement
with Belle's result.

\begin{table}[htb]
\begin{center}
\caption{\label{bellepi0pi0} Fit results for $2 \gamma$ production of
  $\pi^0 \pi^0$ in the low mass energy region by
  Belle~\cite{bellepi0pi0}.}
\begin{tabular}{|l|c|c|c|c|}
\hline 
Parameter & Nominal & $r_{02} = 0$  & No $f_0(Y)$ &  Units \\ \hline
mass($f_0(980)$) &  $982.2\pm1.0$ & $980.2\pm1.0$ &
$983.7^{+1.5}_{-1.0}$ & MeV/c$^2$ \\
$\Gamma_{\gamma \gamma}(f_0(980))$ & $285.5^{+18.2}_{-18.1}$ &
$297.7^{+14.2}_{-13.7}$ & $370.5^{+20.2}_{-18.7}$ & eV \\ \hline
mass($f_0(Y))$ & $1469.7 \pm4.7$ & $1466.8 \pm 0.6$ & -- & MeV/c$^2$ \\
$\Gamma(f_0(Y))$ & $89.7^{+8.1}_{-6.6}$ & $422.4^{+18.4}_{-19.8}$ & -- &
MeV \\
$\Gamma_{\gamma \gamma} B(f_0(Y) \to \pi^0 \pi^0)$ &
$11.2^{+5.0}_{-4.0}$ & $6780.2^{+626.5}_{-574.7}$ & 0 (fixed) & eV \\
\hline
$r_{02}$ & $3.69^{+0.24}_{-0.29}$ & 0 (fixed) & $5.04^{+0.26}_{-0.24}$ &
  \% \\
$B(f_2(1270\to\gamma\gamma)$ & $1.57\pm 0.01$ & $1.62^{+0.02}_{-0.01}$
& $1.52^{+0.13}_{-0.31}$ & $\times 10^{-5}$ \\ \hline
$\chi^2 (ndf)$ & 1010 (615) & 1206 (617) & 1253 (619)  & \\ \hline
\hline
\end{tabular}
\end{center}
\end{table}

\section{BaBar ISR physics}

\begin{figure}[!htb]
\centering
\includegraphics[width=55mm]{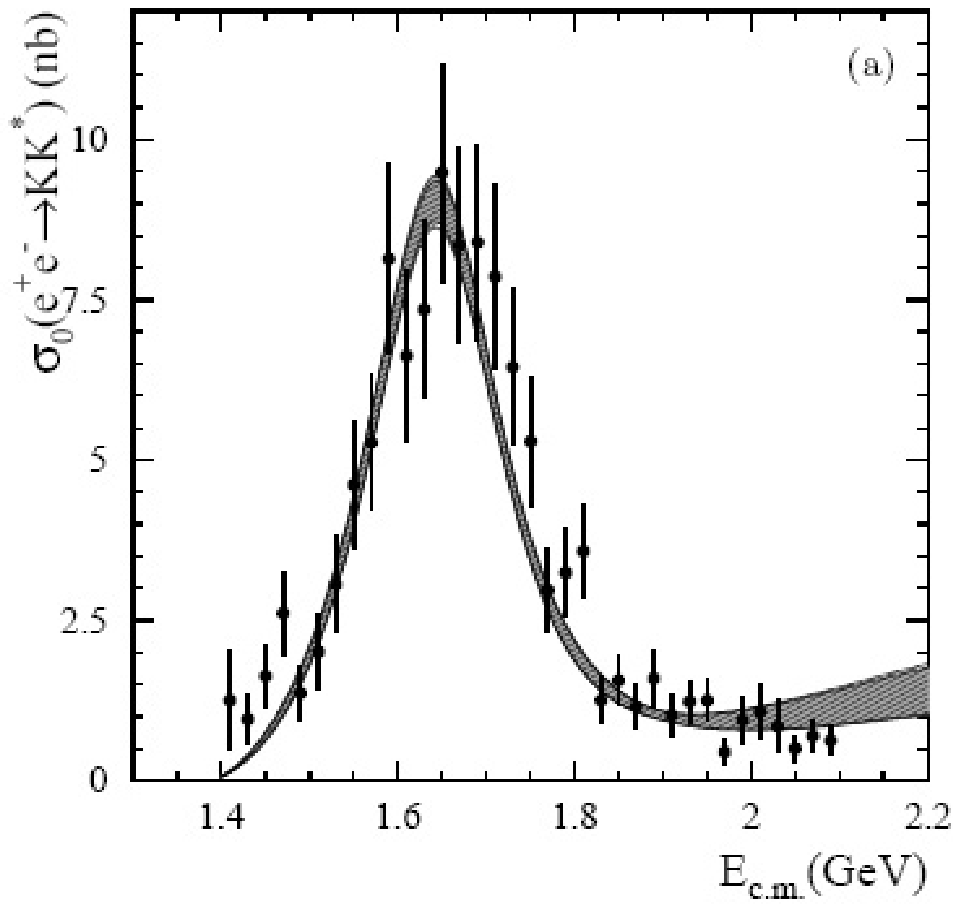}
\includegraphics[width=55mm]{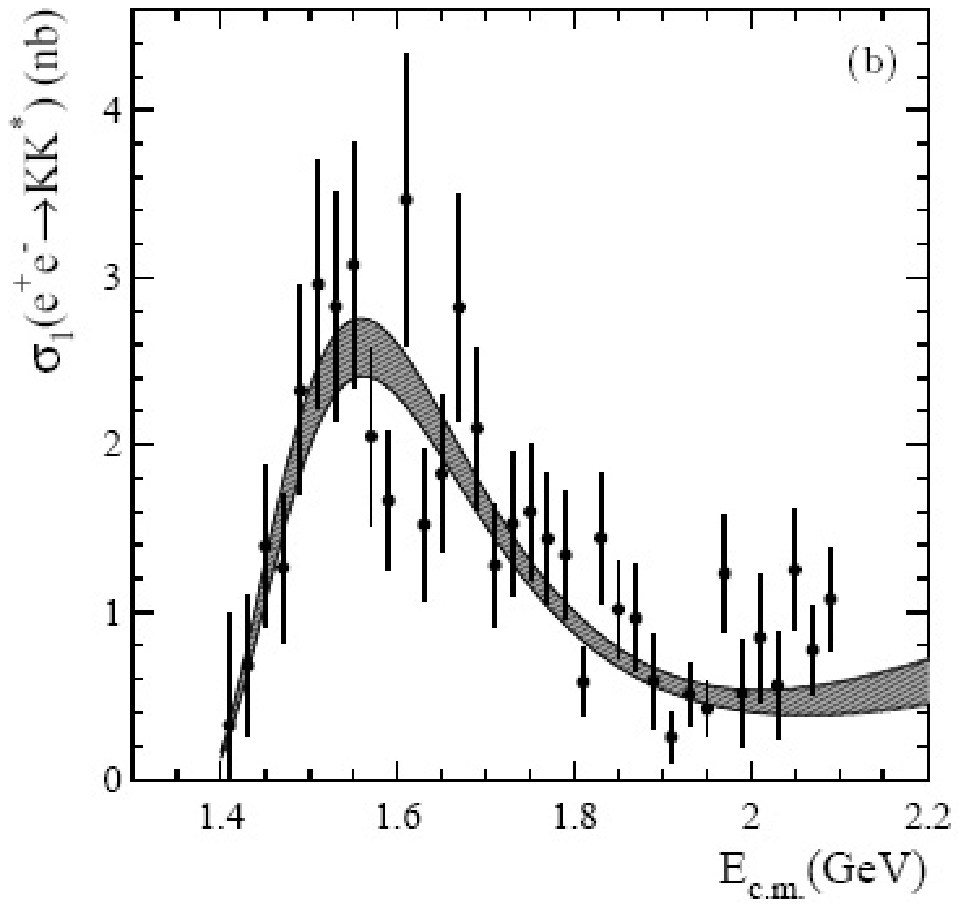}
\caption{The cross section for (a) isoscalar and (b) isovector
  components from a Dalitz plot analysis of ISR $e^+e^- \to KK\pi$ by
  BaBar~\cite{babar_kkpi}. 
}
\label{fig:babar_isr}
\end{figure}

Because of the very high luminosity at B factories, hadron
spectroscopy has also benefited greatly from studies using ISR to
reduce the CM energy below the $\Upsilon(4S)$ to study $e^+e^- \to
hadrons$ from $1 < \sqrt{s} < 5$ GeV.  The BaBar collaboration have used 232 fb$^{-1}$ at
the $\Upsilon(4S)$ to study $e^+e^- \to K^+K^- \pi^0$ and
$K_SK^-\pi^+$ using this technique.  They require that the ISR
$\gamma$ be detected, which forces the hadrons to be within the
fiducial volume of the detector, and fully reconstruct the hadronic
final state.  The Dalitz plots are dominated by $K^*
K^*$ production.  From a Dalitz plot analysis, they have separated the
isoscalar and isovector components and measured their cross sections,
as shown in Fig.~\ref{fig:babar_isr}. The channels are dominated by
resonances that are consistent with the $\phi(1680)$ and
$\rho(1450)$~\cite{babar_kkpi}.

\begin{figure}[!htb]
\centering
\includegraphics[width=58mm]{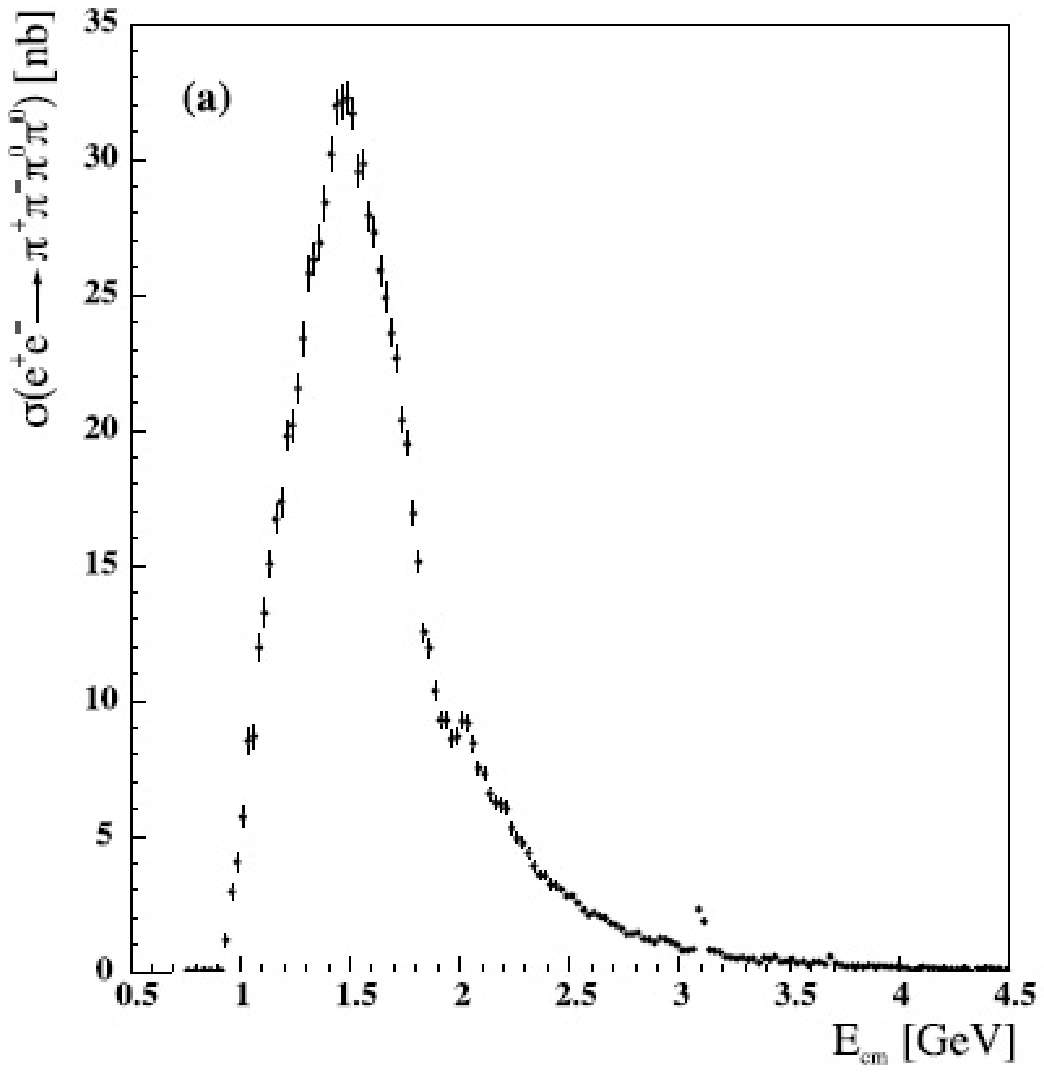}
\includegraphics[width=55mm]{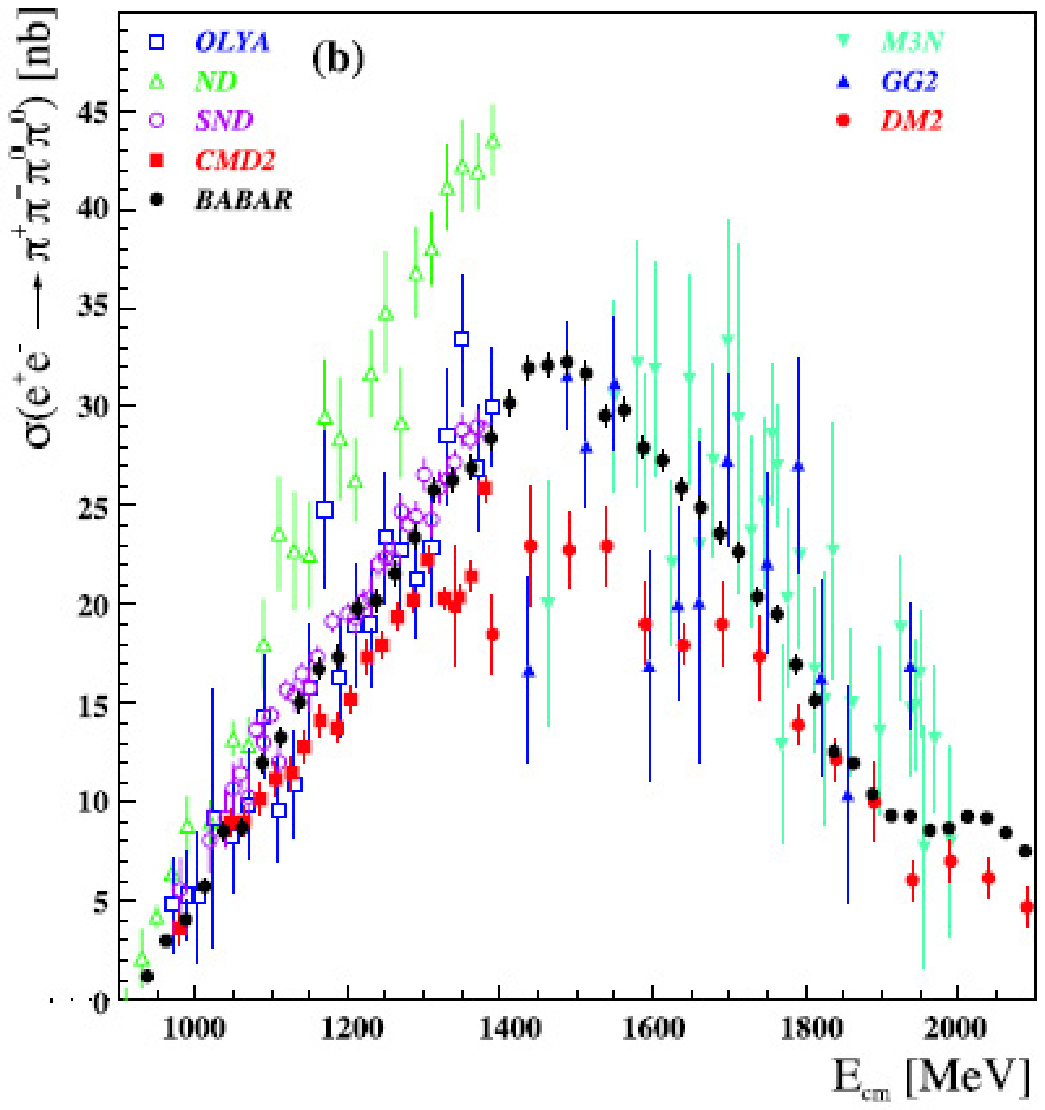}
\caption{Cross section for ISR $e^+ e^- \to\pi^+ \pi^- \pi^0
\pi^0$~\cite{babar_4pi}.  (a) Full energy range. (b) Low energy range
with results from previous experiments.  Results are preliminary.  
}
\label{fig:babar_4pi}
\end{figure}

BaBar has also studied $e^+ e^- \to \pi^+ \pi^- \pi^0
\pi^0$ using ISR events~\cite{babar_4pi}.  Channels with four pions dominate the cross sections in
the $1 < E_{CM} < 2$ GeV region, and are very important for
determinations of the anomalous magnetic moment, $\alpha_{\mu}$, and
the fine structure constant evaluated at the $Z$-pole,
$\alpha(m_Z^2)$.  The cross section, shown in
Figs.~\ref{fig:babar_4pi} a and b, is consistent with SND at low energy and is
a huge improvement about 1.4 GeV, as shown in Fig.~\ref{fig:babar_4pi}
b.  The preliminary precision is 8\%, and it is hoped that it will
reach 5\% over the peak region, which will help improve the precision
of $\alpha_{\mu}$.  This method will be used to improve the precision
of $R$ values at low energy, which Marco Verzocchi pointed out at this
conference could be a bottleneck to future tests of Electroweak physics.

\section{BEPCII and BESIII}
Finally I report briefly the status of BEPCII and BESIII.
BEPCII is a two-ring $e^+e^-$ collider that will run in the tau-charm
energy region ($E_{CM} = 2.0 - 4.2$ GeV, but possibly as high as 4.6
GeV) with a design luminosity of $1\times 10^{33}$ cm$^{-2}$s$^{-1}$
at a beam energy of 1.89 GeV, an improvement of a factor of 100 in
luminosity with
respect to the BEPC. This is accomplished mainly by using
multi-bunches and micro-beta.

\begin{figure}  \centering
   \includegraphics*[width=0.44\textwidth]{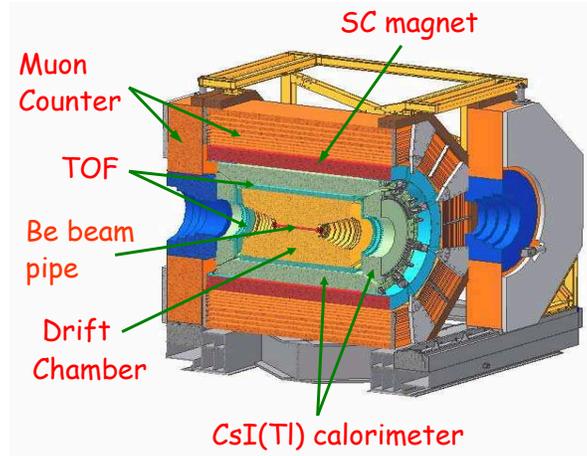}  
  \caption{\label{schematic}Schematic view of the BESIII detector.
    }
 \end{figure}

The BESIII detector consists of a beryllium beam pipe, a helium-based
small-celled drift chamber, Time-Of-Flight counters (TOF) for particle
identification, a CsI(Tl) crystal calorimeter, a super-conducting
solenoidal magnet with a field of 1 Tesla, and a muon identifier using
the magnet yoke interleaved with Resistive Plate
Counters. Fig.~\ref{schematic} shows the schematic view of the BESIII
detector, including both the barrel and end cap portions.
    
\begin{figure}  \centering
   \includegraphics*[width=0.48\textwidth]{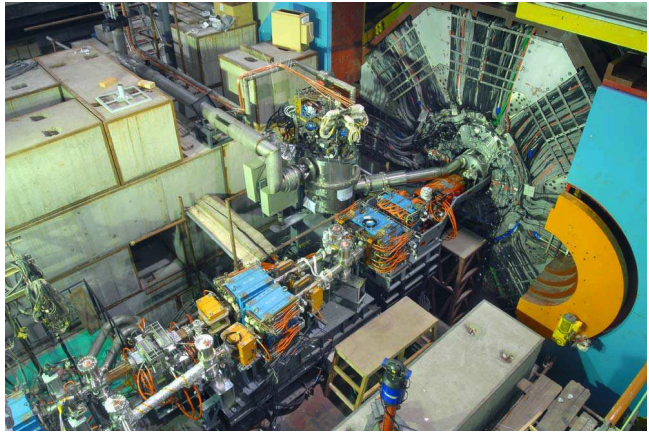}  
  \caption{\label{besatIP}BESIII detector at the IP in June
  2008. Shown are a superconducting quadrupole, the two beam lines,
  and the BESIII detector with the magnet iron open and end caps
  exposed.  }
 \end{figure}

The detector moved to the IP in the spring of this year and is
shown in Fig.~\ref{besatIP} at its final location in June 2008 with
all beam magnets and vacuum pipes in place.
Commissioning of the detector and collider together began in July, and
the first hadronic event was obtained on July 19, 2008. Currently data
at the $\psi(2S)$ is being taken for calibration purposes.

Clearly BESIII with a luminosity of $1 \times 10^{33}$
cm$^{-2}$ s$^{-1}$ will contribute greatly to precision flavor
physics; $V_{cd}$ and $V_{cs}$ will be measured with a statistical
accuracy of better than 1.0\%. $D^0 \bar{D^0}$ mixing will be studied,
and CP violation will be searched for.  Huge $J/\psi$ and $\psi(2S)$
samples will be obtained.  The $\eta_c$, $\chi_{cJ}$, and $h_c$ can be
studied with high statistics.   The high statistics will allow searches for
physics beyond the standard model.  The future is very bright.
More detail on BEPCII and BESIII may be found in Ref.~\cite{fah}.

\begin{acknowledgments}
The author wishes to thank the CLEOc, Belle, and BaBar collaborations
for their help in preparing this paper and especially wants to thank
Ryan Mitchell, Selina Li, Chengping Shen,and Xiaoyan Shen for their
help.

Work supported by Department of Energy contract DE-FG02-04ER41291.
\end{acknowledgments}


\end{document}